\documentclass[twocolumn,preprintnumbers,amsmath,amssymb,showkeys]{revtex4}

\usepackage{graphicx}
\usepackage{dcolumn}
\usepackage{bm}

\begin{document}

\preprint{APS/123-QED}

\title{Critical mass of bacterial populations in a generalized
Keller-Segel model.\\ Analogy with the Chandrasekhar limiting mass of
white dwarf stars}

\author{Pierre-Henri Chavanis and Cl\'ement Sire}
\email{chavanis@irsamc.ups-tlse.fr}
\affiliation{Laboratoire de Physique Th\'eorique - IRSAMC, CNRS
Universit\'e Paul Sabatier, 31062 Toulouse, France }%

\date{\today}

\begin{abstract}
We point out a remarkable analogy between the limiting mass of
relativistic white dwarf stars (Chandrasekhar's limit) and the
critical mass of bacterial populations in a generalized Keller-Segel
model of chemotaxis [Chavanis \& Sire, PRE, {\bf 69}, 016116
(2004)]. This model is based on generalized stochastic processes
leading to the Tsallis statistics. The equilibrium states correspond
to polytropic configurations similar to gaseous polytropes in
astrophysics. For the critical index $n_{3}=d/(d-2)$ (where $d\ge 2$ is the
dimension of space), the theory of polytropes leads to a unique value
of the mass $M_{c}$ that we interpret as a limiting mass. In $d=3$, we
find $M_{c}=202.8956...$ and in $d=2$, we recover the well-known
result $M_{c}=8\pi$ (in suitable units). For $M<M_{c}$, the system
evaporates (in an infinite domain) or tends to an equilibrium state
(for box-confined configurations). For $M>M_{c}$, the system collapses
and forms a Dirac peak containing a mass $M_{c}$ surrounded by a
halo. This paper exposes the model and shows, by simple
considerations, the origin of the critical mass. A detailed
description of the critical dynamics of the generalized Keller-Segel
model will be given in a forthcoming paper.
\end{abstract}

\maketitle

\section{Introduction}
\label{sec_introduction}

Recently, there has been a growing interest for the dynamics and
thermodynamics of systems with long-range interactions
\cite{houches}. Such systems are numerous in Nature and share
fascinating analogies. For example, the statistical
mechanics of stellar systems (like globular clusters and elliptical
galaxies) in astrophysics and the statistical mechanics of large-scale
vortices (like Jupiter's great red spot) in two-dimensional turbulence
present deep similarities \cite{chavhouches} despite the very
different physical nature of these systems. Some connections have also
been developed between the dynamics of stellar systems and the
dynamics of the Hamiltonian Mean Field (HMF) model \cite{yama,cvb} and
of the free electron laser \cite{barre}. More recently, the authors
have studied a model of self-gravitating Brownian particles
\cite{crs,sc,virial1} and discovered striking
analogies with the Bose-Einstein condensation in the canonical
ensemble \cite{bose} and with the phenomenon of chemotaxis in biology
\cite{mass}. In this paper, we point out a novel analogy between
the Chandrasekhar limiting mass of white dwarf stars
\cite{chandramass} and the critical mass of bacterial populations
experiencing chemotactic aggregation.

The name chemotaxis refers to the motion of organisms induced by
chemical signals \cite{murray}. In some cases, the biological
organisms (bacteria, amoebae, endothelial cells, ants...) secrete a
substance (pheromone, smell, food, ...) that has a long-range
attractive effect on the organisms themselves. Therefore, in addition
to their diffusive motion, they move preferentially along the gradient
of concentration of the chemical they secrete (chemotactic flux). When
attraction prevails over diffusion, the chemotaxis can trigger a
self-accelerating process until a point at which aggregation takes
place. This is the case for the slime mold {\it Dictyostelium
discoideum} and for the bacteria {\it Escherichia coli}. This is
referred to as chemotactic collapse. A model of slime mold aggregation
has been introduced by Keller \& Segel
\cite{ks} in the form of two coupled differential equations. A
simplified version of this model has been extensively studied in the
case where the degradation of the secreted chemical can be neglected
and an immediate production is assumed \cite{jl}. In that case, the
Keller-Segel (KS) model become isomorphic to the Smoluchowski-Poisson
(SP) system describing self-gravitating Brownian particles
\cite{sc}. The steady states correspond to isothermal distributions
similar to isothermal stars in astrophysics \cite{chandra}. On the
other hand, the KS model and the SP system can be viewed as standard
mean field Fokker-Planck equations associated with the Boltzmann free
energy \cite{risken,hb2}. In this sense, they are based on ordinary
thermodynamics. Since these systems are strongly dissipative, the
correct statistical ensemble is the canonical ensemble \cite{sc}.

Recently, modified forms of Keller-Segel models have been introduced
in order to describe more general situations
\cite{ph,gamba,lang,filbet,degrad,logotropes,kinbio}. For example, in
Ref. \cite{lang}, we have introduced and studied a generalized
Keller-Segel (GKS) model of chemotaxis taking into account anomalous
diffusion. In this model, the coefficient of diffusion is assumed to
depend on the density like $D(\rho)=D\rho^{\gamma-1}$ with
$\gamma=1+1/n$ (the ordinary KS model is recovered for $\gamma=1$ or
$n\rightarrow +\infty$). Anomalous diffusion is known to appear in
many problems of biology
\cite{murray} and it is likely that it can play a role in the process
of chemotaxis.  The generalized Keller-Segel (GKS) model is isomorphic
to the generalized Smoluchowski-Poisson (GSP) system describing
self-gravitating Langevin particles \cite{lang}. The steady states
correspond to polytropic distributions similar to polytropic stars in
astrophysics \cite{chandra}. On the other hand, the GKS model and the
GSP system can be viewed as nonlinear mean field Fokker-Planck
equations \cite{gen,frank} associated with the Tsallis free energy
\cite{tsallis}. In this sense, they are related to a notion of
(effective) generalized thermodynamics.

For the GKS model, there exists a particular polytropic index
$n_{3}=d/(d-2)$ (where $d\ge 2$ is the dimension of space) at which the
dynamics is critical. For this index, the theory of polytropes leads
to a unique value of the mass $M_{c}$, independent on the size of the
object, that we interpret as a limiting mass. This is the counterpart
of the Chandrasekhar limiting mass
\cite{chandramass} for white dwarf stars which are equivalent, in the
ultra-relativistic limit, to polytropes of index $n=3$ (in $d=3$). This
unexpected analogy between two systems that have apparently nothing in
common was pointed out in \cite{massD} and is here systematically
developed (see also Ref. \cite{critique}).

In this paper, we expose the GKS model (using the notations of
biology) and show its relation to Tsallis generalized
thermodynamics. Then, we show by simple considerations, the origin of
the critical mass $M_{c}$ appearing at the polytropic index
$n_{3}=d/(d-2)$.  For $d=3$, we find that $M_{c}=202.89561...$ in a
proper system of units. For $d=2$ where $n_{3}\rightarrow +\infty$, we
recover the critical mass $M_{c}=8\pi$ corresponding to the ordinary 
KS model in two dimensions whose critical dynamics has been
extensively studied in \cite{sc,virial1,mass} (see
\cite{cp,childress,nagai,herrero,biler1,dolbeault,biler2} for
many rigorous results obtained by applied mathematicians).  For
$d=3$, a detailed description of the critical dynamics of the GKS
model for $M<M_{c}$ and $M>M_{c}$ in bounded and unbounded domains
will be given in a forthcoming paper
\cite{critique}.

\section{The ordinary Keller-Segel model: Boltzmann thermodynamics}
\label{sec_b}

The primitive Keller-Segel model \cite{ks} describing the chemotaxis of bacterial populations \footnote{To be specific, we assume that the biological organisms experiencing chemotaxis are bacteria, but our results can be valid for other  systems like amoebae, cells, social insects... The Keller-Segel model generically describes a wide variety of biological systems. } consists in two coupled differential equations 
\begin{eqnarray}
\label{b1}
\frac{\partial\rho}{\partial t}=\nabla\cdot \left (D_{2}\nabla\rho)-\nabla\cdot (D_{1}\nabla c\right ),
\end{eqnarray}
\begin{eqnarray}
\label{b2}
\frac{\partial c}{\partial t}=D_{c}\Delta c+h(c)\rho-k(c)c, 
\end{eqnarray}
which  govern the evolution of the density of bacteria $\rho({\bf r},t)$
and the evolution of the secreted chemical $c({\bf r},t)$. The
bacteria diffuse with a diffusion coefficient $D_{2}$ and they also
move in a direction of a positive gradient of the chemical
(chemotactic drift). The coefficient $D_{1}$ is a measure of the
strength of the influence of the chemical gradient on the flow of
bacteria. On the other hand, the chemical is produced by the bacteria
with a rate $h(c)$ and is degraded with a rate $k(c)$. It also
diffuses with a diffusion coefficient $D_{c}$. In the general
Keller-Segel model, $D_{1}=D_{1}(\rho,c)$ and $D_{2}=D_{2}(\rho,c)$
can both depend on the concentration of the bacteria and of the
chemical. This can take into account microscopic constraints, like
close-packing effects or anomalous diffusion \cite{kinbio}.

A simplified version of the Keller-Segel model is provided by
the system of equations
\begin{eqnarray}
\label{b3}
\frac{\partial\rho}{\partial t}=\nabla\cdot \left (D\nabla\rho-\chi\rho\nabla c\right ),
\end{eqnarray}
\begin{eqnarray}
\label{b4}
\frac{\partial c}{\partial t}=D_{c}\Delta c+h\rho-kc,
\end{eqnarray}
where the parameters are positive constants.  Another simplification
is obtained in a limit of large diffusivity of the chemical
$D_{c}\rightarrow +\infty$ and for sufficiently large concentrations
of the bacteria (see \cite{jl} and Appendix C of \cite{kinbio} for
details). In that case, Eq. (\ref{b4}) is replaced by a Poisson equation
\begin{eqnarray}
\label{b5}
\Delta c=-\lambda \rho,
\end{eqnarray}
where $\lambda=h/D_{c}$. Equation (\ref{b3}) can be viewed as a
mean-field Fokker-Planck equation associated with a Langevin dynamics
of the form
\begin{eqnarray}
\label{b6}
\frac{d{\bf r}}{dt}=\chi\nabla c+\sqrt{2D}{\bf R}(t),
\end{eqnarray}
where ${\bf R}(t)$ is a white noise, $D$ is a diffusion coefficient and
$\chi$ plays the role of a mobility. The Langevin equation describes a
point organism performing a random walk biased in the direction of a
drift velocity proportional to the local chemical gradient.

The KS model (\ref{b3})-(\ref{b5})
monotonically decreases  the Lyapunov functional
\begin{eqnarray}
\label{b7}
F=-\frac{1}{2}\int \rho c \, d{\bf r}+\frac{D}{\chi}\int \rho\ln\rho
\, d{\bf r},
\end{eqnarray} 
which is similar to a free energy $F=E-T_{eff}S$ where $E=-(1/2)\int
\rho c d{\bf r}$ is the energy of interaction and $S=-\int \rho\ln\rho
d{\bf r}$ is the Boltzmann entropic functional. We have defined an
effective temperature
\begin{eqnarray}
\label{b8}
T_{eff}=\frac{D}{\chi},
\end{eqnarray} 
which is given by a form of Einstein's formula. Using the KS model, we
find that 
\begin{eqnarray}
\label{b8add}
\dot F=-\int\frac{1}{\chi\rho}(D\nabla\rho-\chi\rho\nabla c)^{2}d{\bf r}\le 0,
\end{eqnarray} 
which is similar to the proper version of the $H$-theorem in the
canonical ensemble \cite{gen,hb2}. The stationary solutions of
Eq. (\ref{b3}), corresponding to $\dot F=0$, are given by
\begin{eqnarray}
\label{b9}
\rho=Ae^{\frac{\chi}{D}c},
\end{eqnarray} 
where $A$ is determined by the mass $M$.  This is similar to the
Boltzmann distribution for a system in a potential $-c({\bf r})$ at
temperature $T_{eff}=D/\chi$. The steady states of Eq. (\ref{b3}) are
critical points of the free energy $F$ at fixed mass $M$. They cancel
the first order variations $\delta F-\alpha\delta M=0$ where $\alpha$
is a Lagrange multiplier. Moreover, it can be shown \cite{gen} that
they are linearly dynamically stable for the KS model if and only if
they are (local) {\it minima} of $F$ at fixed mass. The equilibrium
state is obtained by coupling Eq. (\ref{b9}) to Eq. (\ref{b5}) leading
to the Boltzmann-Poisson equation.

These analogies with thermodynamics take even more sense if we remark
that the Keller-Segel model is isomorphic to the Smoluchowski-Poisson
system 
\begin{equation}
\frac{\partial\rho}{\partial t}=\nabla\cdot \left\lbrack
\frac{1}{\xi}\left (\frac{k_{B}T}{m}\nabla\rho+\rho\nabla\Phi\right
)\right\rbrack,
\label{b10}
\end{equation}
\begin{equation}
\Delta\Phi=S_{d}G\rho,
\label{b11}
\end{equation}
describing the dynamics of self-gravitating Brownian particles in an
overdamped limit and in a mean field approximation \cite{crs,hb2}. We have
the correspondence
\begin{equation}
\label{b12}D=k_{B}T/\xi m,\quad  \chi=1/\xi,\quad  c=-\Phi, \quad \lambda=S_{d}G.
\end{equation}
We note, in particular, that the role of the gravitational potential
$\Phi$ in the SP system is played, in the KS model, by the
concentration $-c$ of the chemical. Therefore, the pheromonal
substance has a long-range attractive effect similar to the
gravitational attraction in astrophysics. In biology, the chemotactic
aggregation is mediated by a physical field (the chemical produced by
the organisms) while the nature of the gravitational force in
astrophysics is more abstract.  The equilibrium states of the SP
system correspond to a condition of hydrostatic balance
\begin{equation}
\nabla P+\rho\nabla\Phi={\bf 0}, \qquad \Delta\Phi=S_{d}G\rho,
\label{b13}
\end{equation}
with an isothermal equation of state
\begin{equation}
P=\rho k_{B}T/m.
\label{b14}
\end{equation}
Therefore, the equilibrium states of the KS model and SP system have
the same structure as isothermal stars in astrophysics
\cite{chandra}. Of course, the dynamics of the KS model and SP system
is different from the dynamics of stars which is rather described by
hydrodynamic equations like the Euler-Poisson system \cite{bt}.

\section{The generalized Keller-Segel model: Tsallis thermodynamics}
\label{sec_t}

In a previous paper \cite{lang} (see also \cite{logotropes}), we have
introduced and studied a generalized Keller-Segel model of the form
\begin{eqnarray}
\label{t1}
\frac{\partial\rho}{\partial t}=\nabla\cdot \left
(D\nabla\rho^{\gamma}-\chi \rho\nabla c \right ),
\end{eqnarray}
\begin{equation}
\label{t2}\Delta c=-\lambda \rho.
\end{equation}
Comparing with the primitive Keller-Segel model (\ref{b1})-(\ref{b2}),
it corresponds to $D_{2}=D\gamma\rho^{\gamma-1}$ and $D_{1}=\chi
\rho$. For simplicity, we have considered that the chemical is
determined by a Poisson equation (\ref{t2}). More generally, we can
consider that it satisfies the field equation (\ref{b4}). However, the
analytical results that we obtain in Secs. \ref{sec_m} and \ref{sec_d}
are only valid for the Poisson equation (\ref{t2}).  For $\gamma=1$,
we recover the ordinary Keller-Segel model (\ref{b3})-(\ref{b5}). Equation
(\ref{t1}) can be viewed as a nonlinear mean field Fokker-Planck
equation of the form considered in
\cite{plastino,bukman,gen,frank}. It can be obtained from the stochastic process
\begin{eqnarray}
\label{t3}
\frac{d{\bf r}}{dt}=\chi\nabla c+\sqrt{2D}\rho^{(\gamma-1)/2}{\bf R}(t),
\end{eqnarray}
where ${\bf R}(t)$ is a white noise \footnote{There are other methods
to justify the GKS model (\ref{t1})-(\ref{t2}). It can be obtained
from the master equation, assuming that the probabilities of
transition explicitly depend on the occupation number (concentration)
of the initial and arrival states \cite{kaniadakis,degrad}. It can
also be obtained in a strong friction limit of a generalized kinetic
model taking into account inertial effects \cite{lemou,kinbio}.}.
This equation describes a situation where the mobility $\chi$ is
constant but the diffusion coefficient $D(\rho)=D\rho^{\gamma-1}$ can
depend on the density. This can account for anomalous diffusion and
non-ergodic behaviors.  For $\gamma=1$, we recover the ordinary
Langevin equation (\ref{b6}) with constant diffusion coefficient $D$
and constant mobility $\chi$. The stochastic process (\ref{t3}) has
been introduced by Borland \cite{borland} in relation with Tsallis
generalized thermodynamics \cite{tsallis}. For $\gamma=1$, we have a
pure random walk. In that case the sizes of the random kicks are
uniform and do not depend on where the particle happens to be. For
$\gamma\neq 1$, the size of the random kicks changes, depending on the
distribution of the particles around the ``test'' particle. A particle
which is in a region that is weakly populated [small $\rho({\bf
r},t)$] will tend to have smaller kicks if $\gamma>1$ and larger kicks
if $\gamma<1$. Since the microscopics depends on the actual density,
this creates a bias in the ergodic behavior of the system.

The GKS model decreases the Lyapunov functional
\begin{equation}
F=-\frac{1}{2}\int\rho c d{\bf r}+\frac{D}{\chi}\frac{1}{\gamma
-1}\int (\rho^{\gamma}-\rho) d{\bf r}.
\label{t4}
\end{equation}
It can be interpreted as a generalized free energy of the form
$F=E-T_{eff} S$ where $E=-(1/2)\int \rho c d{\bf r}$ is the energy and
$S=-1/(\gamma-1)\int (\rho^{\gamma}-\rho)d{\bf r}$ is the Tsallis
entropy (the polytropic index $\gamma$ plays the role of the Tsallis
$q$ parameter). The effective temperature $T_{eff}$ is still given by the
Einstein-like formula (\ref{b8}). Using the GKS model, we
find that 
\begin{eqnarray}
\label{t4add}
\dot F=-\int\frac{1}{\chi\rho}(D\nabla\rho^{\gamma}-\chi\rho\nabla
c)^{2}d{\bf r}\le 0,
\end{eqnarray} 
which is similar to the proper version of the $H$-theorem in the
canonical ensemble in a generalized thermodynamical framework
\cite{gen,hb2}. The stationary solution of Eq. (\ref{t1}),
corresponding to $\dot F=0$, is given by
\begin{equation}
\rho=\left \lbrack
\mu+\frac{\chi}{D}\frac{\gamma-1}{\gamma}c\right\rbrack^{1/(\gamma-1)},
\label{t5}
\end{equation}
where $\mu$ is determined by the mass $M$. This corresponds to the
Tsallis distribution with inverse temperature $\beta=1/T_{eff}=\chi/D$
and ``$q$-parameter'' $\gamma$. The steady states of Eq. (\ref{t1})
are critical points of the free energy $F$ at fixed mass $M$. They
cancel the first order variations $\delta F-\alpha\delta M=0$ where
$\alpha$ is a Lagrange multiplier. Moreover, it can be shown that they
are linearly dynamically stable for the GKS model if and only if they
are (local) {\it minima} of $F$ at fixed mass \cite{gen}.  The
equilibrium state is obtained by coupling Eq. (\ref{t5}) to
Eq. (\ref{t2}) leading to the ``Tsallis-Poisson'' equation.

These analogies with generalized thermodynamics take even more sense
if we remark that the GKS model is isomorphic to the
generalized Smoluchowski-Poisson system \cite{lang}:
\begin{equation}
\frac{\partial\rho}{\partial t}=\nabla\cdot \left\lbrack
\frac{1}{\xi}\left (K\nabla \rho^{\gamma}+\rho\nabla\Phi\right
)\right\rbrack,
\label{t6}
\end{equation}
\begin{equation}
\Delta\Phi=S_{d}G\rho,
\label{t7}
\end{equation}
provided that we set
\begin{equation}
\label{t8}D=K/\xi,\quad  \chi=1/\xi,\quad  c=-\Phi, \quad \lambda=S_{d}G.
\end{equation}
In particular, the equilibrium states correspond to a condition of
hydrostatic balance
\begin{equation}
\nabla P+\rho\nabla\Phi={\bf 0}, \qquad \Delta\Phi=S_{d}G\rho,
\label{t9}
\end{equation}
with a polytropic equation of state
\begin{equation}
P=K\rho^{\gamma}.
\label{t10}
\end{equation}
Therefore, the equilibrium states of the GKS model and GSP system have
the same structure as polytropic stars in astrophysics \cite{chandra}. As in
astrophysics, we define the polytropic index $n$ by
\begin{equation}
\gamma=1+\frac{1}{n}.
\label{t11}
\end{equation}

\section{The critical mass of bacterial populations for $n=n_{3}$}
\label{sec_m}

The steady states of the GKS model are determined by substituting the
polytropic distribution (\ref{t5}) in the Poisson equation
(\ref{t2}). The resulting configurations have the same structure as
polytropic stars in astrophysics. Therefore, we can readily apply the
theory of polytropes developed long ago by Emden \cite{emden} to the biological
problem. For sake of generality, we extend these results to a space of
$d$ dimensions \cite{lang}. The dimension $d=2$ (which is important in biology)
will be considered specifically in Sec. \ref{sec_d}.

We consider spherically symmetric configurations. Defining
\begin{equation}
\rho=\rho_{0}\theta^{n}, \qquad \xi=\left\lbrack
\frac{\lambda\chi\rho_{0}^{1-1/n}}{D(1+n)}\right\rbrack^{1/2}r,
\label{m1}
\end{equation}
where $\rho_{0}$ is the central density, and using the Poisson
equation (\ref{t2}) with the steady distribution (\ref{t5}), we find
after simple algebra that the function $\theta(\xi)$ is solution of
the Lane-Emden equation
\begin{equation}
\frac{1}{\xi^{d-1}}\frac{d}{d\xi}\left
(\xi^{d-1}\frac{d\theta}{d\xi}\right )=-\theta^{n},
\label{m2}
\end{equation}
with $\theta=1$ and $\theta'=0$ at $\xi=0$. In the following, we
consider $d>2$ and $n> 0$. It is shown in \cite{lang} that
polytropic configurations are self-confined iff
\begin{equation}
n<n_{5}=\frac{d+2}{d-2}.
\label{m3}
\end{equation}
In that case, the function $\theta(\xi)$ vanishes at a finite distance
$\xi=\xi_{1}$. Consequently, the density $\rho(r)$ vanishes at a
distance $R_{*}$ defining the radius $R_{*}$ of the polytrope.  The
relation between the radius and the central density is
\begin{equation}
\xi_{1}=\left\lbrack
\frac{\lambda\chi\rho_{0}^{1-1/n}}{D(1+n)}\right\rbrack^{1/2}R_{*}.
\label{m4}
\end{equation}
The mass $M=\int_{0}^{R_{*}}\rho S_{d}r^{d-1}dr$ of the configuration
is given by
\begin{equation}
M=S_{d}\rho_{0}\left\lbrack \frac{D(1+n)}{\lambda\chi\rho_{0}^{1-1/n}}
\right\rbrack^{d/2}\int_{0}^{\xi_{1}}\theta^{n}\xi^{d-1}d\xi.
\label{m5}
\end{equation}
Using the Lane-Emden equation (\ref{m2}), we get
\begin{equation}
M=-S_{d}\rho_{0}\left\lbrack
\frac{D(1+n)}{\lambda\chi\rho_{0}^{1-1/n}}
\right\rbrack^{d/2}\xi_{1}^{d-1}\theta'_{1}.
\label{m6}
\end{equation}
Expressing the central density in terms of the radius, using
Eq. (\ref{m4}), and introducing the index
\begin{equation}
n_{3}=\frac{d}{d-2}, 
\label{m7}
\end{equation}
we obtain the mass-radius relation
\begin{eqnarray}
M^{(n-1)/n}R_{*}^{\lbrack (d-2)(n_3-n)\rbrack/n}=\frac{D(1+n)}{\lambda
\chi S_{d}^{(1-n)/n}}\omega_{n}^{(n-1)/n},
\label{m8}
\end{eqnarray}
where
\begin{eqnarray}
\omega_{n}=-\xi_{1}^{(n+1)/(n-1)}\theta'_{1}.
\label{m9}
\end{eqnarray}
This is nothing but the usual mass-radius relation for polytropes
\cite{emden} extended to $d$ dimensions \cite{lang}, and written with
the notations of biology. 

\begin{figure}[htbp]
\centerline{ \includegraphics[width=8cm,angle=0]{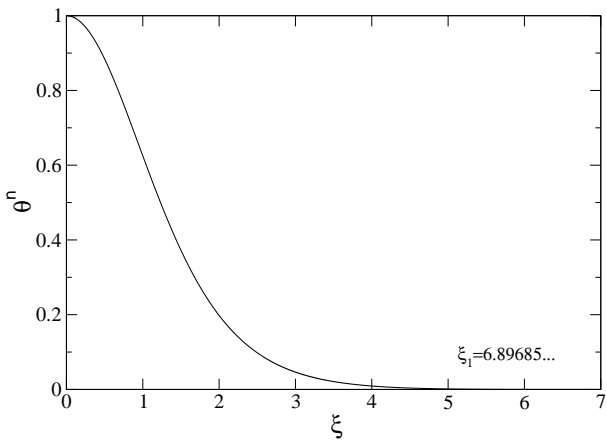} }
\caption[]{Invariant density profile for the 3D generalized
Keller-Segel model at the critical index $n=3$ and critical mass
$M_{c}$.  }
\label{polytrope}
\end{figure}

For $n<n_{3}$ there is one, and only one, steady state for each mass
$M$ and it is stable (global minimum of $F[\rho]$ at fixed mass $M$). Its
radius $R_{*}$ is determined by Eq. (\ref{m8}). For $n_5>n>n_{3}$
there is one, and only one, steady state for each mass $M$ but it is
unstable (saddle point of $F[\rho]$ at fixed mass). The index $n_{3}$
is {\it critical} \cite{lang}. For $n=n_{3}$, steady state solutions
exist for a unique value of the mass
\begin{eqnarray}
M_{c}=S_{d}\left\lbrack
\frac{D(1+n_{3})}{\lambda \chi}\right\rbrack^{n_{3}/(n_{3}-1)}\omega_{n_{3}}.
\label{m10}
\end{eqnarray}
Their radius $R_{*}$ is arbitrary and they are marginally
stable. This yields a family of density profiles of the form 
\begin{eqnarray}
\rho(r)=\rho_{0}\theta^{n_{3}}(\xi_{1}r/R_{*}),
\label{m10add}
\end{eqnarray}
where $\rho_{0}$ is related to $R_{*}$ by Eq. (\ref{m4}). They all
have the same mass (\ref{m10}) and it can be shown that their
free energy is independent of $R_{*}$ (see Appendix F of
\cite{lang}). The invariant profile $\theta(\xi)^{n_{3}}$ is plotted
in Fig. \ref{polytrope}. In $d=3$ where $n_{3}=3$, the critical mass is
\begin{eqnarray}
M_{c}=32\pi\omega_{3}\left (
\frac{D}{\lambda \chi}\right )^{3/2}.
\label{m11}
\end{eqnarray} 
It is found by solving numerically the Lane-Emden equation that
$\omega_{3}=2.01824...$ \cite{chandra}. Therefore, we obtain more
quantitatively
\begin{eqnarray}
M_{c}=202.8956... \left (
\frac{D}{\lambda \chi}\right )^{3/2}.
\label{m12}
\end{eqnarray} 
It can be convenient to introduce dimensionless variables or,
equivalently, to take $D=\lambda=\chi=1$. In that case, the only
control parameter is the mass $M$. With this system of units, the
critical mass in $d=3$ is $M_c=202.8956...$.

The critical mass (\ref{m11}) of bacterial populations is the
counterpart of the Chandrasekhar limiting mass for white dwarf stars
\cite{chandramass}. The analogy stems from the fact that, in the
ultra-relativistic limit, white dwarf stars are equivalent to
polytropes with the critical index $n=3$ (in $d=3$). Then, applying
the theory of polytropes, Chandrasekhar \cite{chandramass} obtains a
unique value of the mass
\begin{eqnarray}
M_{Chandra}=\left (\frac{3}{32\pi^{2}}\right )^{1/2}\omega_{3} \left
(\frac{hc}{G}\right )^{3/2}\frac{1}{(\mu H)^{2}},
\label{m13}
\end{eqnarray} 
where $h$ is the Planck constant, $c$ is the velocity of light, $G$ is
the constant of gravity and $H$ is the mass of the hydrogen atom ($\mu$
is the molecular weight). In terms of the solar mass, it reads
\begin{equation}
M_{Chandra}=0.196701...\biggl ({hc\over G}\biggr )^{3/2}{1\over (\mu
H)^{2}}\simeq 5.76 M_{\odot}/\mu^{2}.
\label{m14}
\end{equation}
In his more general treatment of partially relativistic white dwarf
stars, Chandrasekhar \cite{partially} shows that the mass (\ref{m14})
represents a limit above which there is no equilibrium state
\footnote{The structure and the stability of (relativistic) white dwarf stars in $d$ dimensions has been studied in \cite{massD}. It is found that the dimension $d=3$ of our universe plays a very particular (marginal) role.}. In that case, the system is
expected to collapse leading ultimately to a neutron star or a black
hole.

\section{The two-dimensional case}
\label{sec_d}

In two dimensions ($d=2$), the critical polytropic index
$n_{3}\rightarrow +\infty$. In that case, $\gamma=1$ and we recover
the usual Keller-Segel model (\ref{b3})-(\ref{b5}) corresponding to
normal diffusion. The stationary solution is obtained by substituting
the Boltzmann distribution (\ref{b9}) in the Poisson equation
(\ref{b5}). The resulting configurations have the same structure as
isothermal stars in astrophysics. Their structure in $d$ dimensions
has been described in \cite{sc}.

\begin{figure}[htbp]
\centerline{ \includegraphics[width=8cm,angle=0]{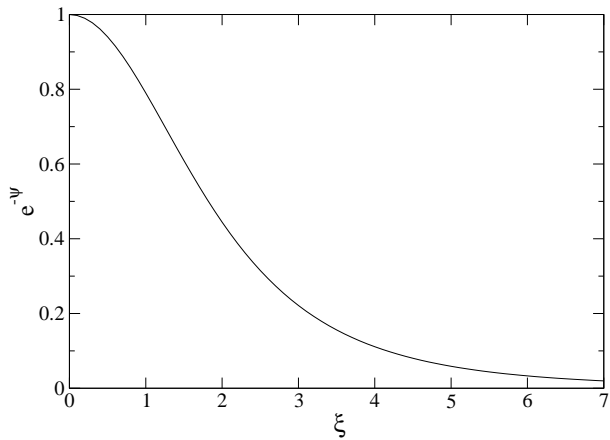} }
\caption[]{Invariant density profile for the two-dimensional
Keller-Segel model at the critical mass $M_{c}$.  }
\label{isothermal}
\end{figure}

We consider spherically symmetric configurations. Defining
\begin{equation}
\rho=\rho_{0}e^{-\psi}, \qquad \xi=(\lambda\chi \rho_{0}/D)^{1/2}r,
\label{d1}
\end{equation}
where $\rho_{0}$ is the central density, we find after simple algebra
that $\psi$ is solution of the Emden equation
\begin{equation}
\frac{1}{\xi^{d-1}}\frac{d}{d\xi}\left
(\xi^{d-1}\frac{d\psi}{d\xi}\right )=e^{-\psi},
\label{d2}
\end{equation}
with $\psi=0$ and $\psi'=0$ at $\xi=0$. For $d=2$, the density profile
extends to infinity but the total mass is finite. The mass
$M=2\pi\int_{0}^{+\infty}\rho r dr$ is given by
\begin{equation}
M=\frac{2\pi D}{\chi\lambda}\int_{0}^{+\infty} e^{-\psi}\xi d\xi.
\label{d3}
\end{equation}
Using the Emden equation (\ref{d2}), we get
\begin{equation}
M=\frac{2\pi D}{\chi \lambda}\lim_{\xi\rightarrow +\infty} \xi \psi'(\xi). 
\label{d4}
\end{equation}
In $d=2$, the Emden function is known analytically \cite{sc}:
\begin{equation}
e^{-\psi}=\frac{1}{\left (1+\frac{1}{8}\xi^{2}\right )^{2}},
\label{d5}
\end{equation}
and we find that $\xi\psi'\rightarrow 4$ for $\xi\rightarrow
+\infty$. This leads to a unique value of the mass
\begin{eqnarray}
M_{c}=\frac{8\pi D}{\chi\lambda}.
\label{d6}
\end{eqnarray}
Therefore, unbounded steady states of the KS model in two dimensions
exist for a unique value of the mass (\ref{d6}) and they are
marginally stable \cite{mass}. This yields a family of density
profiles of the form
\begin{eqnarray}
\rho(r)=\frac{\rho_{0}}{\left
(1+\frac{\lambda\chi\rho_{0}}{8D}r^{2}\right )^{2}},
\label{m10add2}
\end{eqnarray}
that are parametrized by the central density $\rho_{0}$. They all
have the same mass (\ref{d6}) and it can be shown that their
free energy is independent of $\rho_{0}$ \cite{mass}. The invariant
profile $e^{-\psi(\xi)}$ is plotted in Fig. \ref{isothermal}.

The critical mass of bacterial
populations (\ref{d6}) is the counterpart of the critical mass or
critical temperature
\begin{eqnarray}
M_{c}=\frac{4k_{B}T}{Gm}, \qquad k_{B}T_{c}=\frac{GMm}{4},
\label{d6b}
\end{eqnarray}
of isothermal spheres in 2D gravity (see Ref. \cite{mass} for a
description of this analogy). The existence of a critical  mass, or
critical temperature, for systems described by the 2D Boltzmann-Poisson
system is known for a very long time in astrophysics
\cite{ostriker,stodolkiewicz,salsberg,klb,paddy,at,ap,sc,new} and in
the statistical mechanics of 2D point vortices
\cite{jm,lp,caglioti,chavhouches}. It has been rediscovered in the
context of chemotaxis in
\cite{cp,childress,nagai,herrero,biler1,dolbeault,biler2,mass}.
Comparing Eq. (\ref{d4}) with Eq. (\ref{m6}) we find that for $d=2$
and $n=n_{3}\rightarrow +\infty$, we have the limit
\begin{eqnarray}
\lim_{d\rightarrow 2} n_{3}\omega_{n_{3}}=4.
\label{d7}
\end{eqnarray}
This relation can also be obtained from Eq. (79) in \cite{lang}. With
this relation (\ref{d7}), we can obtain the critical mass (\ref{d6}) of
the KS model in two dimensions as a particular case of the critical
mass (\ref{m10}) of the GKS model when $d\rightarrow 2$. In the context of generalized thermodynamics, this corresponds to the limit $\gamma\rightarrow 1$ where we pass from the Tsallis (polytrope) distribution to the Boltzmann (isothermal) distribution. In the present context, this corresponds to the passage from anomalous ($\gamma>1$) to normal ($\gamma=1$) diffusion. 

\section{Conclusion}
\label{sec_c}

Our study tends to suggest that a notion of generalized thermodynamics
can be useful in the context of chemotaxis when the biological
organisms experience anomalous diffusion. Therefore, the chemotactic
aggregation of bacterial populations could be an important {\it
physical} system where ideas of generalized thermodynamics apply. The
generalized Keller-Segel model (\ref{t1})-(\ref{t2}) that we have
introduced is related to the Tsallis form of entropic functional. This
is a very rich model because {\it it combines both elements of generalized
thermodynamics and long-range interactions}. Therefore, it enters in
the general class of nonlinear mean field Fokker-Planck equations
introduced by Chavanis \cite{gen}. In fact, we can
develop an effective generalized formalism for an even larger class of
entropic functionals than the Tsallis entropy. For example, we can
consider a generalized stochastic process 
\begin{equation}
\frac{d{\bf r}}{dt}=\chi\nabla c+\sqrt{2 \chi P(\rho)\over\rho}{\bf R}(t),\label{c2rr}
\end{equation}
where the diffusion
coefficient is given by $D(\rho)=\chi P(\rho)/\rho$ where $P(\rho)$ is an
almost arbitrary function which plays the role of an equation of state
in the analogy with barotropic stars in astrophysics
\cite{lang}. Then, we get the GKS model
\begin{eqnarray}
\label{c1}
\frac{\partial\rho}{\partial t}=\nabla\cdot \left \lbrack \chi \left
(\nabla P-\rho\nabla c \right )\right\rbrack,
\end{eqnarray}
\begin{equation}
\label{c2}\Delta c=-\lambda \rho.
\end{equation}
The equilibrium states correspond to the  condition of hydrostatic
equilibrium (\ref{b13}). This system satisfies a formalism of generalized
thermodynamics \cite{kinbio} for a class of entropic functionals
\begin{eqnarray}
\label{c1ch}
F=-\frac{1}{2}\int\rho c d{\bf r}+\int\rho\int^{\rho}\frac{P(\rho')}{\rho^{'2}}d\rho'd{\bf r},
\end{eqnarray}
larger than the
Tsallis entropic functional \footnote{Note that the model
(\ref{c1})-(\ref{c2}) is still a particular case of the primitive
Keller-Segel model (\ref{b1})-(\ref{b2}). However, the primitive
Keller-Segel model does not satisfy a formalism of generalized
thermodynamics. Apparently, the most general form of Keller-Segel
model satisfying a formalism of effective generalized thermodynamics
(effective temperature, Einstein's relation, $H$-theorem,...)  is
provided by Eqs. (134)-(135) of \cite{kinbio}.}.  However, the Tsallis
thermodynamics is convenient to describe deviations from the Boltzmann
thermodynamics in a simple manner. This leads to models that are still
analytically tractable. For example, the GKS model
(\ref{t1})-(\ref{t2}) can be studied in great detail
\cite{lang,logotropes,critique}.

For the index $n_{3}$, the GKS model (\ref{t1})-(\ref{t2}) presents a
critical dynamics involving a ``universal'' mass $M_{c}$ {\it
independent on the size of the system}. In astrophysics, this result
is well-known in $d=3$ and is connected to the limiting mass of white
dwarf stars discovered by Chandrasekhar \cite{chandramass}. We can
immediately transpose this result to the biological context leading to
the critical mass (\ref{m12}) of bacterial populations. This result
was implicit in \cite{lang} and it has been developed in the present
paper with emphasis. Thus, our paper reveals the analogy of the
existence of a critical mass, found initially for white dwarf stars,
to exist also for bacterial populations driven by chemotaxis. Now, the
question concerns the detailed description of the critical dynamics of
the GKS model at $n=n_{3}$. This will be reported in a forthcoming
paper
\cite{critique}. From the present study, we anticipate that the
critical dynamics of the GKS model at $n_{3}=3$ in $d=3$ will be very
similar to the critical dynamics of the ordinary KS model with
$n_{3}=+\infty$ in $d=2$ studied in \cite{sc,virial1,mass}. This
analogy will be confirmed in \cite{critique}. For $M<M_{c}$, we find
that the system evaporates (in an infinite domain) or tends to an
equilibrium state (in a finite domain) corresponding to an {\it
incomplete polytrope} confined by the box. For $M>M_{c}$, we find that
the system collapses. In a finite time, it forms a Dirac peak
containing a mass $M_{c}$ surrounded by a halo that has a pseudo
self-similar evolution.  These results are similar to those found in
$d=2$. In conclusion, we can interpret the mass $M_{c}$ as {\it a
limiting mass above which the system undergoes chemotactic
collapse}. This strengthens the analogy with the Chandrasekhar mass of
white dwarf stars. There is, however, a great conceptual difference
between the two. Indeed, the Chandrasekhar mass is defined in terms of
fundamental constants so it has a universal value. By contrast,
the critical mass of bacterial populations depends on the parameters
$\lambda$, $\chi$ and $D$ that are not universal and that change from
experiment to experiment. Furthermore, the index $n=3$ in $d=3$ is
special in astrophysics because it corresponds to the index of
ultra-relativistic and completely degenerate white dwarf stars
\cite{chandra}. In biology, there is {\it a priori} no reason why the
index $n=n_{3}$ should be selected in the dynamics of
bacterial populations (except in $d=2$ where it corresponds to a
situation of ordinary diffusion).

\end{document}